\documentclass[multphys,vecarrow]{svmult}

\usepackage{graphicx}
\usepackage{amssymb}
\usepackage{amsmath}
\usepackage{subeqnar}
\usepackage{epsfig}
\usepackage{citesort}

\begin{document}
\frontmatter
\mainmatter

\bibliographystyle{unsrt}

\title*{Numerical stress response functions of static granular layers}
\author{A.P.F. Atman and P. Claudin}
\institute{
Laboratoire des Milieux D\'esordonn\'es et H\'et\'erog\`enes -- UMR 7603\\
4, Place Jussieu -- case 86, 75252 Paris Cedex 05, France\\
e-mail: {\tt atman@ccr.jussieu.fr}, {\tt claudin@ccr.jussieu.fr}\\
}
\authorrunning{A.P.F. Atman and P. Claudin}

\date{\today}

\maketitle

\begin{abstract}

We investigate the stress response function of a layer of grains, i.e. the stress profile in response to a
localized overload. The shape of the profile is very sensitive to the packing arrangement, and is thus a good 
signature of the preparation procedure of the layer.
This study has been done by the use of molecular dynamics numerical simulations. Here, for a given rain-like 
preparation, we present the scaling properties of the response function, and in particular the influence of
the thickness of the layer, and the importance of the location of the overload and measurement points
(at the boundaries, in the bulk).
\end{abstract}


The statics of granular materials has been a rich field of research over the last few years. One of the reasons 
of this interest is that a system of grains reaches its mechanical equilibrium after the particles are `jammed' 
in configurations in relation to the previous dynamics of the grains. As a consequence, the distribution of the 
stresses in a static piling of grains depends, in a subtle manner, on the preparation procedure of the system. 
A now famous example is that of the sand pile for which the pressure profile at the bottom is different when 
built from a hopper, or by successive horizontal layers (a `rain' of grains). In the former situation, this 
profile exhibits a clear `dip' below the apex of the pile \cite{smid81}, whereas it shows a flat `hump' in the 
latter \cite{vanel99}.

\begin{figure}[t]
\centerline{\epsfig{file=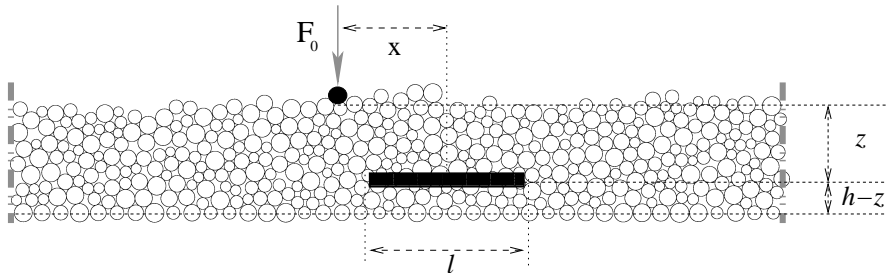,height=3.6cm,angle=0}}
\caption{System geometry and notations. The stress measure is done at a position $(x,z)$ from the applied force
$F_0$ and at a distance $h-z$ from the bottom of the layer. To compute the stress, contact forces are integrated
over a linear extension $\ell$. Note that we use horizontal periodic boundary conditions.
\label{fig:scheme}}
\end{figure}

In fact, more interesting than the case of a pile, is the study of the stress response function of a layer of 
grains, i.e. the stress profiles in response to a localized overload, see figure \ref{fig:scheme}. This geometry
allows much more variety in the preparation of the layer. Experiments have been performed for instance with
compacted, loose or sheared  layers \cite{serero01,geng01,geng03}, whose response pressure profiles are different
enough from each other to be a kind of `signature' of their texture. Besides, such a response test is the
elementary `brick' with which other situations can be deduced, e.g. that of the pile \cite{reydellet03}. It
is then also well adapted for the comparison of the different models of stress distribution in granular media.

Our aim is to perform extensive simulations of assemblies of grains, in order to provide precise two-dimensional
numerical data of stress response functions. The control of all the parameters of the simulations, as well as
the ability of measuring both micro (grain size) and macro (system size) quantities, ensure a useful and
interesting feed back to the experiments and the models.

Although our ambition is to be as general and systematic as possible, we shall in this paper, present only few 
results concerning some scaling and size effects of this response function. In particular we shall restrict to 
a single (rain-like) preparation history and study the influence of the thickness of the layer, as well as that
of the location of the overload and measure points. Besides, the overloading procedure requires careful and
important validity tests (e.g. linearity, additivity, reversibility) that will be also described below.

\section{The simulations}

\begin{figure}[t]
\begin{center}
\epsfig{file=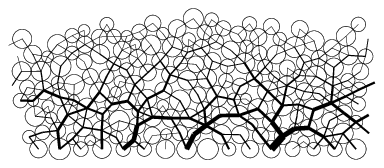,width=0.47\linewidth,angle=0}
\hfill
\epsfig{file=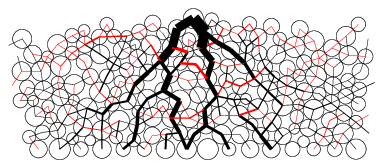,width=0.47\linewidth,angle=0}
\end{center}
\caption{Example of the granular layers produced by our simulations.
Left: Force chain network after the deposition under gravity of $N=1600$ grains. This picture corresponds
to the equilibrium configuration after 628100 MD steps.
Right: Response due to an overload localized over a single grain after subtraction of the initial forces.
The amplitude of this overload corresponds
to the weight of five average grains. This picture has been obtained after 29500 additional MD steps to reach
a new equilibrium configuration. In black, contact forces have been increased in response to the overload,
whereas in gray they have been decreased.
\label{fig:echanforce}}
\end{figure}

In figure \ref{fig:echanforce}, we show a part of a typical simulated layer, where we can see the initial
force chain network and the response produced by a localized overload applied to a single grain after the
initial forces have been subtracted. The simulations
are performed using a classical molecular dynamics (MD) algorithm in three successive stages: preparation,
deposition and overloading that are described in the following.

We start with the preparation of $N$ polydisperse grains, with radii homogeneously distributed between $R_{min}$
and $R_{max} = 2 R_{min}$. These $N$ particles are put on the nodes of a grid with aspect ratio $1:4$ -- this
aspect ratio is needed for the appropriate study of the response functions, see below. This lattice ends at its
bottom with a fixed horizontal line of similar particles which will be used as the support for deposition -- the
distances between these bottom grains are small enough to avoid grain evasion.

The deposition stage consists of untying the particles from the grid and, under gravity and horizontal periodic
boundary conditions, letting them evolve with random initial velocities. In this `rain-like' deposition the
grains interact with elastic and friction forces, and the system evolves under classical Newton equations and
molecular dynamics rules (e.g. predictor-corrector, Verlet algorithm). The time to reach the equilibrium
depends on the characteristics of the particles. We chose $\mu = 0.5$ for the contact friction coefficient. The
rheology of the contacts is that of Kelvin-Voigt with $k_n=1000$ N/m and $k_t=750$ N/m for the normal and
horizontal contact stiffness. The viscosity $g_n$ is chosen in order to get a critical damping. At last, the
gravity is set to unity. The equilibrium criteria consist of the five following tests which are applied after
each period of 100 MD time steps: (1) the number of gained/lost contacts during this period has to be zero;
(2) the number of sliding contacts between particles also has to vanish; (3) the integrated force measured at
the bottom of the layer must be equal to the sum of the weight of all the grains; (4) all the particles have to
have, at least, two contacts; and (5) the total kinetic energy has to be lower than some low threshold. Once
these criteria are all satisfied, the deposition is stopped, and the overloading phase can begin.

\begin{figure}[t]
\centerline{\epsfig{file=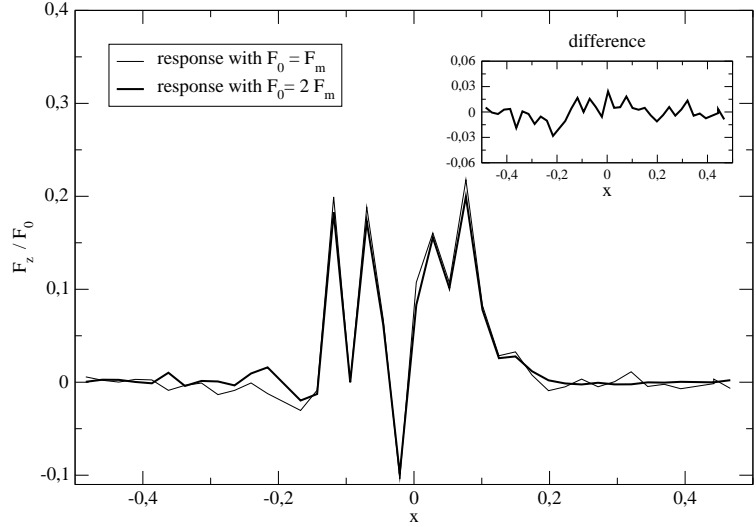,height=10cm,angle=270}}
\caption{In this picture we show one of the tests applied to the layers (linearity, additivity and reversibility)
to settle the optimal value for $F_0$. We present here the linearity test which consists in comparing the force
response profiles with different values of $F_0$. These curves have been obtained on a layer with $N=400$ grains
($h \sim 7 D$), and $F_m = 2.5$ average grain weight. Rescaled by the overload force, the difference of the two
responses is reasonnably small. This difference is due to small slips in the contacts between the grains.
\label{fig:linear}}
\end{figure}

The overload is applied to a single grain with a force $F_0$ that we wish to be small enough to not cause any
rearrangement of the layer structure. The determination of the amplitude of the loading force is then a crucial
part of the simulation. For that purpose we made several tests which check the reversibility, the additivity
and the linearity of the stress response. These tests (see below for more details) let to conclude that the
optimal magnitude of $F_0$ is of the order of few times the weight of an average grain. In the overloading
procedure, $F_0$ is split in 10000 MD time steps, i.e. in each time step the force is increased in 1/10000 of
$F_0$, and it remains constant and equal to $F_0$ after that. Again, the equilibrium criteria are those explicited
above, except that the integrated force at the bottom of the layer has to be equal to the weight of all the
grains \emph{plus} $F_0$.

Only the linearity test is illustrated here, in figure \ref{fig:linear}: it consists in comparing the force
response after loading with different values of $F_0$. We find that the response is satisfactorily
linear in $F_0$ as long as its magnitude does not exceed $\sim 40$ times the average grain weight. With the
other tests we have checked that loading and unloading with the very same $F_0$ gives the original contact
force distribution with a good precision, and that two simultaneous overloads located at two different
positions give a response which is perfectly equivalent to the sum of the two responses computed from the
two corresponding single overloads.

\section{Results and discussion}

In order to measure the vertical stress $\sigma_{zz}$, we integrate the contact forces over a set of grains.
In fact, we do not make use of the usual formula $\sigma_{\alpha \beta} = 1/S \sum_i f^i_\alpha r^i_\beta$,
where the sum is computed over the contacts between the grains in the `volume' $S$ (here we are two-dimensional),
$f_\alpha$ being $\alpha^{th}$ component of the considered contact force, and $r_\beta$ the $\beta^{th}$ component
of the corresponding distance vector between the grains in contact. Rather, we take a horizontal `line' of grains
of length $\ell$. The corresponding stress $\sigma_{zz}$ is then equal to the sum of the vertical components of
the forces carried by the contacts of one of the sides of the line -- say the upper one -- divided by $\ell$. This
stress measure can be done at any depth $z$ ($z=h$ means that the measure is done on the bottom) and centered
at any horizontal distance $x$ from the overload point. As we are particulary interested in $\sigma_{zz}$ profiles
along $x$ at a given $z$, our choice is here better adapted than the usual stress formula which would mix
together grains of (slightly) different depths in $S$. This is particulary important at small $z$. Besides, with
this stress definition, the integral of $\sigma_{zz}$ over $x$ at a fixed $z$ is exactly equal to $F_0$ at any
scale. This property is crucial to normalize and compare data from layers of different thickness.

\begin{figure}[t]
\centerline{\epsfig{file=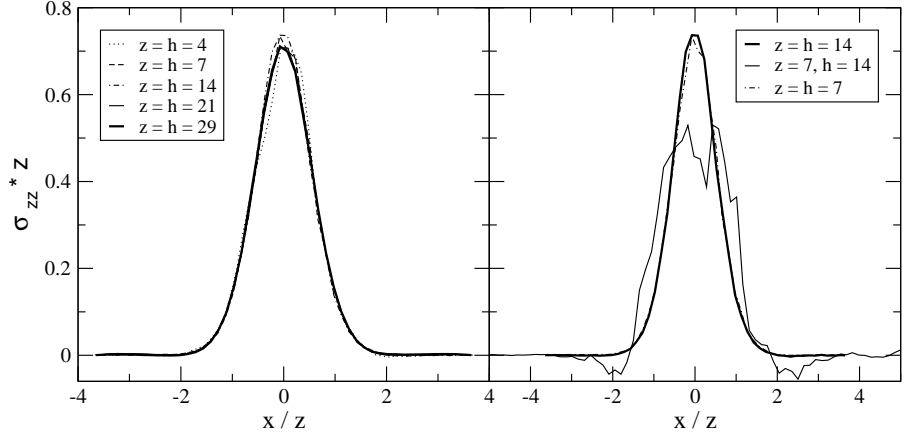,height=12cm,angle=270}}
\caption{Response functions. Several curves are shown for different values of $z$ and $h$. On the left, the
response profiles collapse when measured at the bottom of the layers and rescaled by $h$. On the right, the
response in the bulk. Although more averaging is needed, this graph suggests interesting features for $z<h$.
All these curves have been computed with $\ell = 0.77 z$.
\label{fig:result}}
\end{figure}

The response function at depth $z$ is obtained by making the difference of the stress profile $\sigma_{zz} (x,z)$
measured on the layer with the additional force $F_0$, with a copy of the very same layer without overload. This
difference is ensemble averaged over several overload positions and layer samples. We present here the results
for five different layer sizes, with $N = 100, 400, 1600, 3600$ and $6400$ grains. The $1:4$ aspect ratio gives
layers of average thickness of $4, 7, 14, 21$ and $29$ mean particle diameters $D$, respectively. The  $10, 5,
5, 4$ and $4$ different realizations give in the end $62, 60, 65, 84$ and $86$ different overload points --
on average, there are six grains between two consecutive overload points.

The results are shown on figure \ref{fig:result}. One can see that the normalized profiles measured at the bottom
of layers of different thickness $h$ collapse when they are linearly rescaled by $h$, and show a single peak whose
width is of the order of $h$. This is precisely the reason why it is important to have layers four times wider
than thick -- $2h$ on both sides of the overload point. When the response is computed in the bulk, the structure
of the response profile looks more complicated. The present curve needs however more ensemble averaging and
should be taken as preliminary. These results were all obtained using a measure scale $\ell = 0.77 z$ for all
layer sizes. A systematic study of the importance of this integration measure length $\ell$ is under way. We 
are also working on the calculation of the other stress components $\sigma_{xx}$ and $\sigma_{xz}$.

These scaling studies are currently extended to other preparation histories, and in particular to the cases of
sheared or more anisotropic layers of grains. At last, we plan to test these data against the predictions of
anisotropic elasticity which, depending on the values of the different parameters, can give various stress
response profiles \cite{otto03}.

\section*{Acknowledgements}

We are indebted to G. Combe for a decise help on the MD codes.
We also thank the granular group of the LMDH for stimulating discussions.


\end{document}